\documentclass[a4paper]{article}
\usepackage{geometry}

\usepackage{setspace}
\usepackage{float}
\usepackage{graphicx}
\usepackage{epstopdf}
\usepackage{standalone}
\usepackage{amsmath,amsfonts,amsthm,bm} % Math packages
\usepackage{color}

\usepackage{pdflscape}
\usepackage{authblk}
\usepackage{subcaption}
\usepackage[flushleft]{threeparttable}
\usepackage{enumitem}
\usepackage{xcolor}
\usepackage{soul}
\usepackage{makecell,booktabs}
\usepackage{caption,dcolumn}
\newcolumntype{d}[1]{D{.}{.}{4}}

\usepackage{dsfont}

\usepackage{multirow}
\usepackage{longtable}
\usepackage{array}
\newcolumntype{P}[1]{>{\centering\arraybackslash}p{#1}}

\usepackage [english]{babel}
\usepackage [autostyle, english = american]{csquotes}
\MakeOuterQuote{"}

\newcommand{\mbR}{{\mathbb R}}

\usepackage[numbers,sort&compress]{natbib}
\usepackage[citecolor=blue]{hyperref}

\usepackage{mathtools}

\newcommand{\interior}[1]{%
 {\kern0pt#1}^{\mathrm{o}}%
}
\usepackage [english]{babel}
\usepackage [autostyle, english = american]{csquotes}
\usepackage{csquotes}

\usepackage{algorithm,algpseudocode}
\usepackage{caption}
\usepackage[makeroom]{cancel}

\makeatletter
\newcommand*\bigcdot{\mathpalette\bigcdot@{.5}}
\newcommand*\bigcdot@[2]{\mathbin{\vcenter{\hbox{\scalebox{#2}{$\m@th#1\bullet$}}}}}
\makeatother

\usepackage{tabularx}

\title{\textbf{A Spatio-Temporal Dirichlet Process Mixture Model for Coronavirus Disease-19}}

\author[1,2]{Jaewoo Park}
\author[1]{Seorim Yi}
\author[3]{Won Chang}
\author[4]{Jorge Mateu}

\affil[1]{Department of Statistics and Data Science, Yonsei University}
\affil[2]{Department of Applied Statistics, Yonsei University}
\affil[3]{Division of Statistics and Data Science,  University of Cincinnati}
\affil[4]{Department of Mathematics, University Jaume I}

\begin{document}

\maketitle

\begin{abstract}
Understanding the spatio-temporal patterns of the coronavirus disease 2019 (COVID-19) is essential to construct public health interventions. Spatially referenced data can provide richer opportunities to understand the mechanism of the disease spread compared to the more often encountered aggregated count data. We propose a spatio-temporal Dirichlet process mixture model to analyze confirmed cases of COVID-19 in an urban environment. Our method can detect unobserved cluster centers of the epidemics, and estimate the space-time range of the clusters that are useful to construct a warning system. Furthermore, our model can measure the impact of different types of landmarks in the city, which provides an intuitive explanation of disease spreading sources from different time points. To efficiently capture the temporal dynamics of the disease patterns, we employ a sequential approach that uses the posterior distribution of the parameters for the previous time step as the prior information for the current time step. This approach enables us to incorporate time dependence into our model in a computationally efficient manner without complicating the model structure. We also develop a model assessment by comparing the data with theoretical densities, and outline the goodness-of-fit of our fitted model. 
%We also provide useful visualizations that can inform public health interventions for infectious diseases, such as social distancing.
%diagnose risk 
%effective range of risk boundaries important..
\end{abstract}

%To describe the temporal dynamics, we employ a sequential approach that uses the posterior distribution for the previous time step as the prior information for the current time step; this enables us to incorporate time dependence into our model without complicating the model structure or inflating the computational cost.

\noindent


\section{Introduction}\label{sec:introduction}

The coronavirus disease 2019 (COVID-19) pandemic has led to millions of deaths worldwide and presents an unprecedented challenge to economics, social life, and public health systems. Understanding the dynamics of the disease is essential to plan and implement effective intervention measures. Most of the disease modeling efforts in the literature have been focusing on modeling and predicting the total incidence, prevalence and mortality observed within certain regions or countries \citep[cf.][]{choi2020estimating, fanelli2020analysis, anastassopoulou2020data, barlow2020accurate}. While such analysis is helpful for regional or national agencies to manage and plan resources allocation over time, it may not be informative about how the disease is spread within the study area over time.  

Considering that COVID-19 spreads quickly through close contact with infected people \citep{cdc2020coronaspread}, detecting the disease hot spots and understanding their dynamics over time and space is important for an effective intervention. Indeed, understanding the spatial distribution of disease hot spots, along with their size, duration, and proximity to landmarks can provide insights into disease outbreaks within the local area. This can be done through analyzing geocoded epidemiological data at a fine spatio-temporal scale, which records the geographic location (home addresses) and time of confirmed cases. In this line, we analyze here the spatio-temporal point patterns of COVID-19 cases in the city of Cali (Colombia).

Our data provides every confirmed case’s exact location and time information, offering
vital insights for the spatio-temporal interaction between individuals concerning the disease spread in a metropolis. 
In contrast, usual aggregated data lack precise information about individual cases and
present a significant challenge in modeling the spatio-temporal dynamics of human-to-human disease transmission when capturing the fine spatial heterogeneity of case
distribution in a small region. Aggregated data may lose fine-grained spatio-temporal information, which will lead the administrative officials to make biased decisions.

Our unique high-resolution dataset for individual cases of COVID-19 in Cali, Colombia, highlights the typical behavior of the spread of an infectious disease, that of forming clusters along a region with varying dimensions in space and time. This sort of triggering mechanism can be fully analyzed using point process methodology. Indeed, spatio-temporal point processes provide a natural framework for modeling COVID-19 cases by regarding the  observations as a realization from the process. 

Spatio-temporal analysis of COVID-19 plays a pivotal role in understanding the dynamics of the spread of COVID-19. Just a  few studies attempt to model the dynamics of COVID-19 using point processes. \cite{Gajardo2021PPforCOVID19} propose a point process regression framework of COVID-19 cases and deaths conditioned on mobility and economic covariates. \cite{Giudici2021NetworkPP} focus on country-level case prediction in 27 European countries by augmenting spatio-temporal point process model with mobility network covariates. \cite{ReinformentSTPP} introduce a generative and intensity-free point process model based on an imitation learning framework to track the spread of COVID-19 and forecast county-level cases in the United States. Closer to our point process approach, \cite{Briz-etal2022} propose a mechanistic spatio-temporal point process model by accounting for both human mobility and geographical proximity. \cite{chiang2022hawkes} adapt Hawkes processes \citep{hawkes1971spectra} to explain a contagious disease pattern at an areal level. Although both approaches can describe the characteristics of epidemic outbreaks through the conditional intensity functions, it is challenging to specify the space-time location of disease-spreading centers. Recently, \cite{park2022interaction} developed an interaction Neyman-Scott point process \citep{neyman1952theory} to detect unobserved cluster centers. However, the model cannot explain temporal dependencies. Furthermore, this method is of limited applicability to large samples (e.g., $N=10,000$) because it requires birth-death Markov chain Monte Carlo (MCMC) \citep{moller2003statistical}, which suffers from the slow mixing. This motivates the development of a new approach that allows researchers to study spatio-temporal patterns of the disease within a faster inferential framework. 

In this manuscript, we develop a spatio-temporal point process model that can automatically select the number of disease spreading clusters and their location by adopting a Dirichlet stochastic process into our framework. Dirichlet process Gaussian mixture models have been widely used for various scientific problems \citep{kottas2007bayesian,ji2009spatial,taddy2010autoregressive,reich2011spatial,kottas2012spatial,chang2020regularized}. Examples include clustering for population genetics data \citep{reich2011spatial} and for market segmentation \citep{chang2020regularized}. From estimated space and time range parameters, we can diagnose the degree of COVID-19 risk. In addition, to quantify the impact of the outbreak on major landmarks in the city, we incorporate landmarks range parameters into our model. We can outline the time-varying effect of landmarks, which provide useful epidemiological information for constructing public health interventions.

Our method is computationally practical because the MCMC algorithm can be easily implemented from the full conditional distribution. This is one of the advantages of our method over the Neyman-Scott-based approaches \citep{mrkvivcka2014two,park2022interaction} which require expensive simulations for parameter estimation. Furthermore, we implement a sequential approach that uses the information from the previous time step, which can provide a quick and realistic analysis.

%Furthermore, our fully Bayesian approach allows adaptive updates of model parameters for given prespecified time windows, which can provide a quick and realistic analysis. 

The remainder of this paper is organized as follows. In Section~\ref{sec:data-description}, we describe the different data sources analyzed in this study. In Section~\ref{sec:model}, we propose a spatio-temporal Dirichlet process mixture model and describe Bayesian inference for the model. We introduce our rolling-window strategies with implementation details. We also provide a model assessment strategy to validate our method. In Section~\ref{sec:application}, we apply our methods to COVID-19 datasets in Cali, Colombia. We show that our model can detect the spatio-temporal cluster centers of COVID-19 cases and provide the impact of the clusters across different times. Furthermore, we can quantify the time-varying effect of the landmarks on COVID-19 cases, which can provide important epidemiological interpretations. We conclude with a summary and discussion in Section~\ref{sec:discussion}.

\section{Data: COVID-19 cases in Cali, Colombia}
\label{sec:data-description}

\begin{figure}[!t]
\centering
\includegraphics[width=0.80\linewidth]{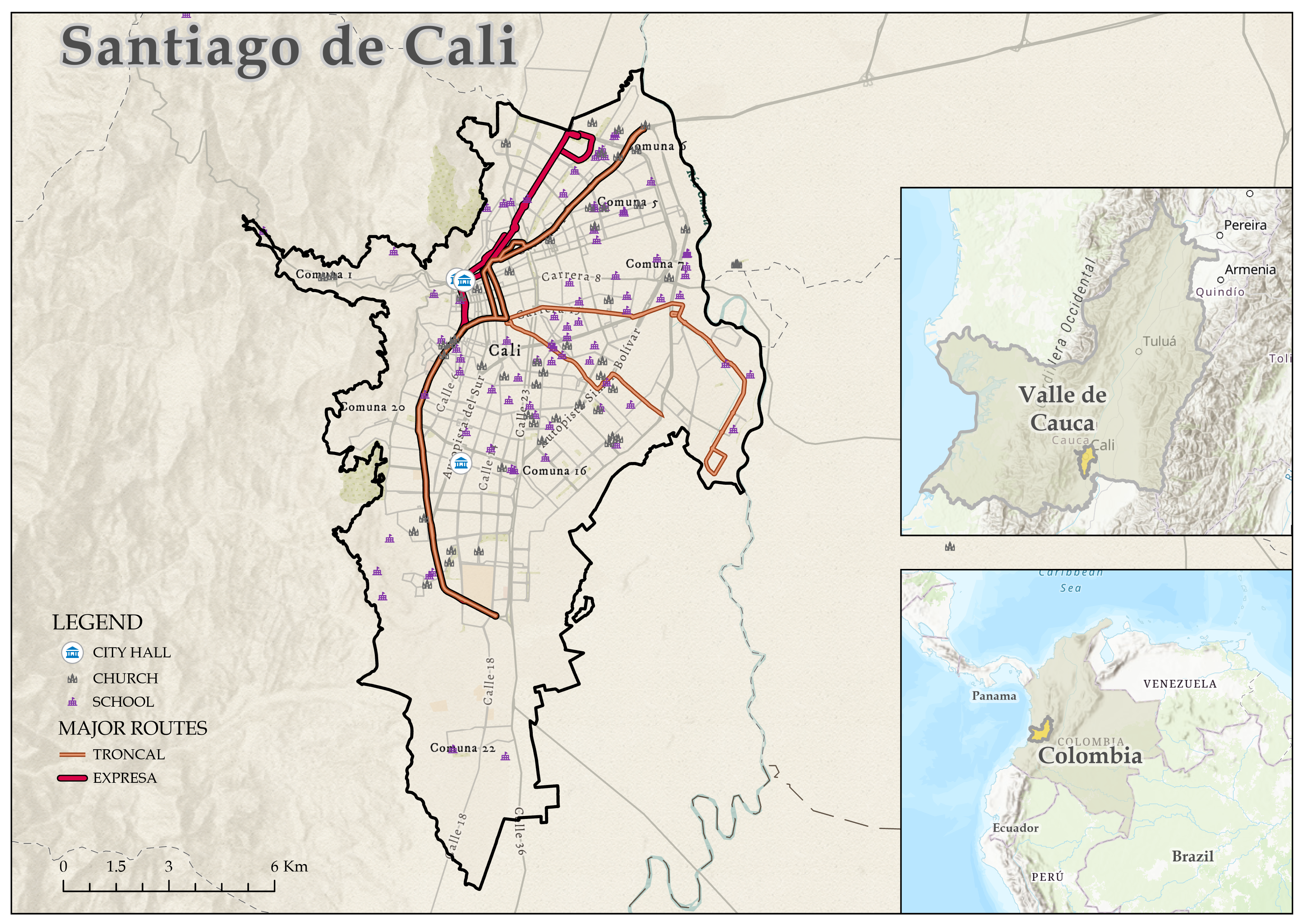}
\caption{Geographical location of Cali}
\label{fig:cali}
\end{figure}

The COVID-19 data that motivates our approach is provided by the Municipal Public Health Secretary of Cali (Colombia)\footnote{\url{https://www.cali.gov.co/salud/}}, institution that documents individual-level confirmed cases of acute respiratory infections due to the new virus COVID-19. This is part of a running program from the National Surveillance System in Public Health (SIVIGILA). Cali represents one of the major cities in Colombia, the capital of Valle del Cauca department, and the most populated city in southwest Colombia, with about 2.2 million residents according to the 2018 census. The city spans 560.3 square kilometers (216.3 square miles) with 120.9 square kilometers (46.7 square miles) of urban area. As the only major Colombian city with access to the Pacific coast, Cali is the leading industrial and economic center in the country's south, with one of Colombia's fastest-growing economies \citep{Cali_intro}. See Fig.~\ref{fig:cali} to depict the location of the city of Cali within Colombia.

Cali is geographically diverse, with an altitude descent from west to east. The city is mainly flat with a small part of mountainous areas on the western city border, leading to a higher population density in the southeastern compared to the northwestern of the city. Owing to its proximity to the equator, there are no significant seasonal variations in Cali \citep{Cali_intro}.

\begin{figure}[!t]
\centering
\begin{subfigure}[b]{0.33\linewidth}
\includegraphics[width=\linewidth, height=2.13in]{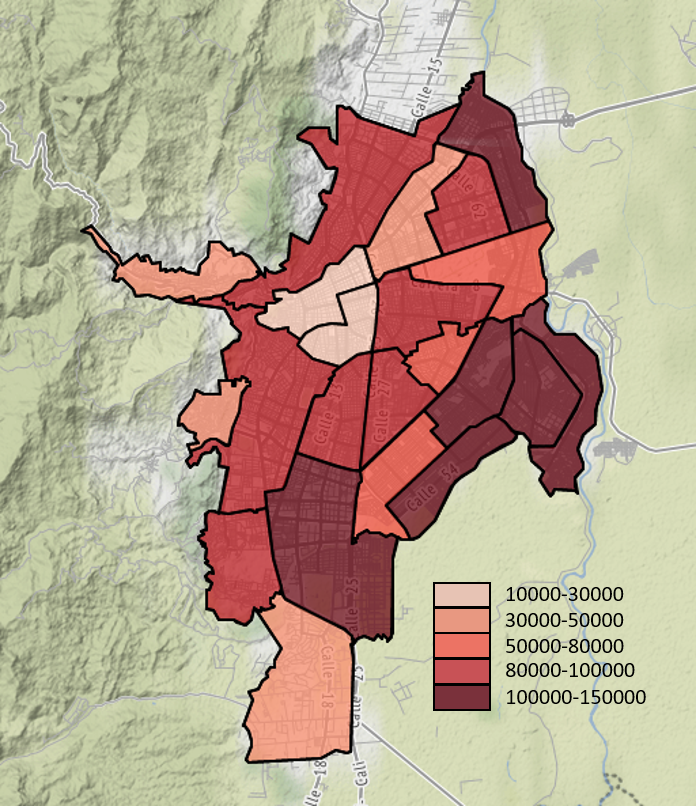}
\caption{Population distribution}
\end{subfigure}
\begin{subfigure}[b]{0.33\linewidth}
\includegraphics[width=\linewidth, height=2.13in]{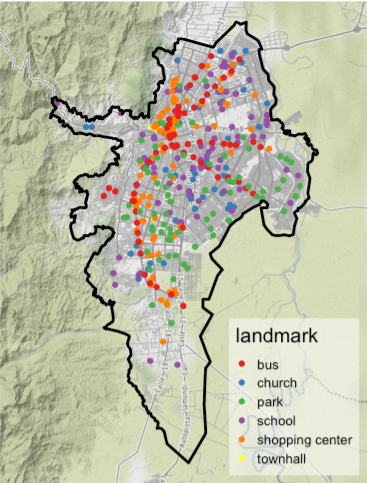}
\caption{Landmarks}
\end{subfigure}
\caption{(a) Population distribution in Cali.
Each polygon bounded by black lines represents a comuna (a municipality-level subdivision in Cali); there are 22 comunas in the city of Cali.
(b) Landmarks in Cali.
Each dot represents the landmark's location, and its color indicates the type of the landmark.}
\label{fig:terrain-and-population}
\end{figure}

As shown in Fig.~\ref{fig:terrain-and-population}(a), more than half of the population concentrates in neighborhoods of low socioeconomic strata located mainly in the east, northeast, and west. Almost a tenth of the population under the line of poverty agglomerates in the city's eastern neighborhoods. The population with higher socioeconomic strata distributes in the other city areas, concentrating the wealthiest population in the city's south.

\begin{figure}[!t]
\centering
\begin{subfigure}[h]{0.24\linewidth}
\includegraphics[width=\linewidth]{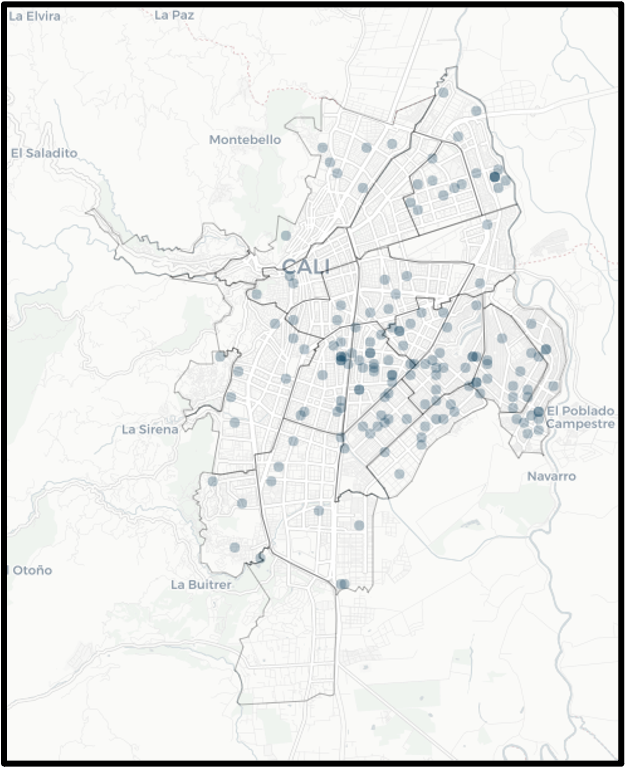}
\caption{March 29, 2020}
\end{subfigure}
\begin{subfigure}[h]{0.24\linewidth}
\includegraphics[width=\linewidth]{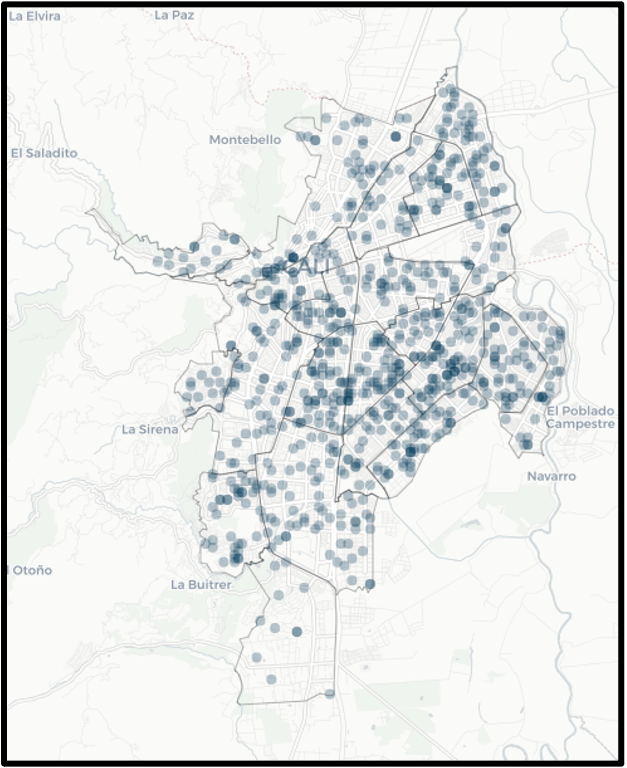}
\caption{May 17, 2020}
\end{subfigure}
\begin{subfigure}[h]{0.24\linewidth}
\includegraphics[width=\linewidth]{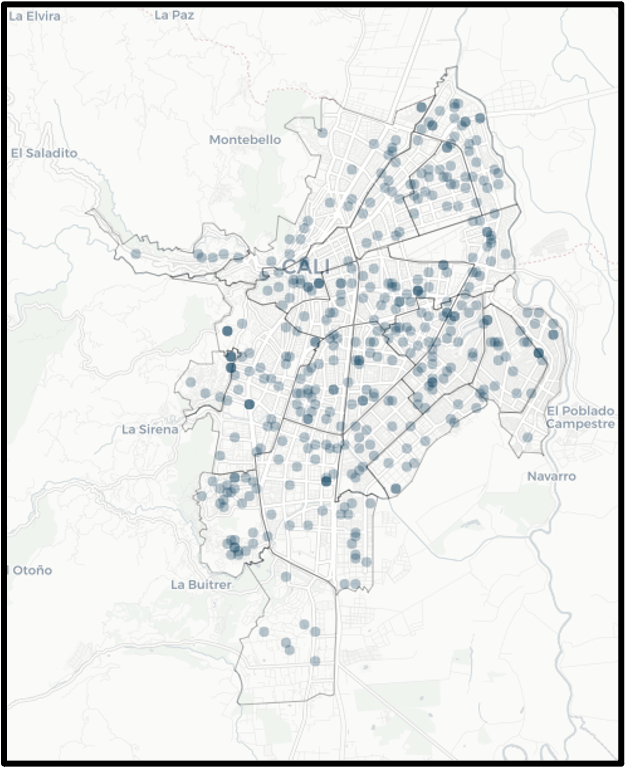}
\caption{July 12, 2020}
\end{subfigure}
\caption{Snapshots of confirmed COVID-19 cases at three particular weeks. Each dot represents the location of a confirmed case. Note that darker dots indicate multiple dots being overlapped.}
\label{fig:cases-snapshots}
\end{figure}

Our dataset consists of 16,309 records of cases from March 02 to July 25 of 2020. Specifically, a COVID-19 case was recorded once confirmed, with the diagnosed date of the patient and the geographical location (measured in longitude and latitude) of their residence. The testing procedures were carried out across the entire urban area, with similar testing rates in each local community. See \cite{Cuartas} for a general description and exploratory analysis of COVID-19 cases in Cali.  Unlike other commonly-seen COVID-19 datasets that only report the aggregated number of cases or deaths at a state or county level, this dataset records the exact location and time information of each single confirmed case, which opens the possibility to look at the data from the perspective of point patterns. To have an overall picture of the evolution, Fig.~\ref{fig:cases-snapshots} presents the spatial distribution of confirmed cases at three particular weeks in Cali.

We note that on March 12, the country declared a state of emergency. Then the authorities announced the mandatory isolation for the entire city for just eight days \citep{COVID19COL}. The first case reported in the city based on people who went to health services occurred in high socioeconomic strata. However, the disease quickly spreads and concentrates in the most vulnerable areas with low socioeconomic strata. After early efforts of the government to contain the pandemic, inevitably, the virus spreads throughout the city, affecting a large part of the population. The above public health decisions are known not significantly to affect the dynamics of the virus spreading, and thus we do not consider the impact of these decisions in our model.

Besides COVID-19 events, we also collected the location of a number of distinct landmarks in Cali, including town halls, churches, schools, shopping centers, bus stations, and parks, from the Administrative Department of Municipal Planning\footnote{\url{https://www.cali.gov.co/planeacion/}}, as these locations play an important role in understanding the wide and rapid spreading of the virus. The landmark dataset has three town halls, 48 small and large churches, 79 schools, 143 shopping centers, 65 bus stations, and 71 parks. Fig.~\ref{fig:terrain-and-population}(b) shows the exact locations of these collected landmarks.

\section{A spatio-temporal Dirichlet process mixture model}\label{sec:model}

In this section, we propose a spatio-temporal Dirichlet process Gaussian mixture model (STDPG) that can capture and describe unobserved spatio-temporal cluster events of COVID-19. We additionally obtain maps of COVID-19 risk.

\subsection{Model framework} \label{sec:model_framework}

Consider the realization of a point pattern $\lbrace (\mathbf{x}_i,t_i)\rbrace_{i=1}^{N}$ from the spatio-temporal point process $\mathbf{X}$, defined over the bounded domain $\mathcal{S}\times \mathcal{T}$. Here, $\mathcal{S} \in \mbR^{2}$, $\mathcal{T} \in \mbR^{+}$ denotes the spatial region of interest and the study period, respectively. We consider latitude and longitude coordinates with a normalized time domain $\mathcal{T} =[0,1]$. In our context, each $(\mathbf{x}_i,t_i)$ represents the location and time of a confirmed individual. Let $\mathbf{c}=\lbrace (\mathbf{c}^{s}_{j},c^{t}_{j})\rbrace_{j=1}^{M} \in \mathcal{S}\times \mathcal{T}$ be the unobserved space-time cluster centers. For the $i$th observation, we introduce the cluster membership variable $g_i \in \lbrace 1,\cdots, M\rbrace$. We model the membership distribution via a Dirichlet process prior \citep{ishwaran2001gibbs} to automatically choose the number of clusters. Especially, we consider the distribution of $g_i$ as
\[
g_i ~\sim \mbox{Categorical}(q_1,\cdots,q_M),
\]
where $q_1,\cdots,q_M$ are cluster probabilities. We use the stick breaking prior \citep{sethuraman1994constructive} for cluster probabilities as 
\begin{equation}
q_j = \begin{cases}
      U_1,  & \mbox{for}~j=1 \\
      U_j\prod_{k=1}^{j-1}(1-U_{k}) & \mbox{for}~j=2,\cdots,M
\end{cases}
\label{DPprior}
\end{equation}
where $U_1,\cdots,U_k \sim_{\mbox{iid}}~\mbox{Beta}(1,b_u)$ with a rate parameter $b_u$. Following \cite{reich2011spatial}, we use an hyperprior for $b_u$ defined as $b_u \sim \mbox{Gamma}(1,1/4)$. In \eqref{DPprior}, the first mixture probability $q_1$ is modeled as a beta random variable $U_1$. For $j=2,\cdots,M$, the subsequent mixture probabilities $q_j$ can be obtained through multiplication between the remaining probability ($1-\sum_{k=1}^{j-1}q_k$) and the proportion assigned to the $j$th cluster component ($U_j$). Although the stick breaking prior considers infinite $M$ in theory, we can approximate it with a reasonably large $M <\infty$ in practice. In this case, the remaining probability $1-\sum_{k=1}^{j-1}q_k$ becomes close to 0 for large $j$ (i.e., no observations come from the $j$th membership). Therefore, the finite representation in \eqref{DPprior} can automatically estimate the number of non-empty space-time clusters $M^{\ast}$ out of $M$ possible clusters. 

We can also incorporate the landmark information into STDPG. Consider we have $l=1,\cdots, p$ different types of landmarks and let $\lbrace \mathbf{z}^{l}_{i} \rbrace_{i=1}^{N} \in \mathcal{S}$ be the location of the closest landmark $l$ from the $i$th observation. Given the cluster membership $g_i$, the space-time locations given by confirmed individuals can be modeled as
\begin{equation}
f( (\mathbf{x}_i,t_i) |g_i=j) \propto \frac{1}{\omega_{s}^{2}\omega_{t}} \exp\Big(-\frac{||\mathbf{x}_i-\mathbf{c}^{s}_{j}||^2}{2\omega_{s}^{2}}-\frac{(t_i-c^{t}_{j})^2}{2\omega_{t}^{2}}\Big)\times \prod_{l=1}^{p}\frac{1}{\omega_{l}} \exp\Big(-\frac{||\mathbf{z}^{l}_{i}-\mathbf{c}^{s}_{j}||^2}{2\omega_{l}^{2}}\Big),
\label{STDPmodel}    
\end{equation}
where $\omega_s,\omega_t,\lbrace \omega_l \rbrace_{l=1}^{p}$ denote space, time, and landmarks range parameters, respectively. In our context, $\omega_s$ and $\omega_t$ control the width of space-time cluster events, and $\omega_l$ can measure the effect of the closest landmark $l$. Smaller $\omega_l$ implies that the $l$th landmark is located nearby space-time cluster centers; therefore, the $l$th landmark can be regarded as important in determining disease hot spots. We set non-informative priors for range parameters and update them sequentially based on results from previous time windows. We describe the details of sequential update procedures in Section~\ref{subsec:window}.

From the Gaussian mixture model described above, the membership variable for each $i$th observation is sampled from 
\[
g_i ~\sim \mbox{Categorical}(\tilde{q}_{i1},\cdots,\tilde{q}_{iM}),
\]
where
\begin{equation}
\tilde{q}_{ij} \propto q_j \frac{1}{\omega_{s}^{2}\omega_{t}} \exp\Big(-\frac{||\mathbf{x}_i-\mathbf{c}^{s}_{j}||^2}{2\omega_{s}^{2}}-\frac{(t_i-c^{t}_{j})^2}{2\omega_{t}^{2}}\Big)\times \prod_{l=1}^{p}\frac{1}{\omega_{l}} \exp\Big(-\frac{||\mathbf{z}^{l}_{i}-\mathbf{c}^{s}_{j}||^2}{2\omega_{l}^{2}}\Big)
\label{membershipprob}
\end{equation}
for $j=1,\cdots,M$. Intuitively, the first term considers the space-time clustering, and the second term measures the impact of the closest landmark on the membership probabilities. Note that the stick breaking prior for $q_j$ in \eqref{DPprior} encourages only $M^{\ast} (< M)$ clusters to be non-empty (i.e., at least a single observation comes from the cluster). From this model specification, the full likelihood function is 
\begin{equation}
L(\bm{\theta}|\mathbf{X}) \propto \prod_{i=1}^{N}\Bigg[
\sum_{\forall j~\mbox{non-empty}} q_j \frac{1}{\omega_{s}^{2}\omega_{t}} \exp\Big(-\frac{||\mathbf{x}_i-\mathbf{c}^{s}_{j}||^2}{2\omega_{s}^{2}}-\frac{(t_i-c^{t}_{j})^2}{2\omega_{t}^{2}}\Big)\times \prod_{l=1}^{p}\frac{1}{\omega_{l}} \exp\Big(-\frac{||\mathbf{z}^{l}_{i}-\mathbf{c}^{s}_{j}||^2}{2\omega_{l}^{2}}\Big)\Bigg],
\label{likelihood}
\end{equation}
where $\bm{\theta} = (\omega_s,\omega_t,\lbrace \omega_l \rbrace_{l=1}^{p})$ is a parameter vector. We provide the conditional distributions for all parameters in the supplementary material. 

\subsection{Rolling-window analysis}\label{subsec:window}

For a spatio-temporal analysis of point processes, the observed events need to be separated into different time steps to capture the dynamics of event occurrence over time. The length of time step needs to be chosen based on domain knowledge, so that the resulting spatio-temporal pattern provides scientifically meaningful representation of the disease spread dynamics. Based on the previous studies of the early stages of the COVID 19 spread, two weeks is the average period of recovery from infection \citep{ki2020epidemiologic, choi2020estimating}; the impact of a confirmed case at a specific space-time location would become weaker as time goes by. Therefore, it would be most informative to utilize the dataset collected over the last several weeks to detect space-time cluster centers of disease spread. 

The rolling-window updates also have practical advantages over all at once updates. Although our model shows much faster mixing than interaction point process models \citep{goldstein2014attraction, park2022interaction}, fitting STDPG with the entire dataset ($N=16,309$) would be still computationally expensive. Instead, we fit STDPG to the subset of the dataset from the considered time window while at the same time incorporating information from the previous time window into the current model fit. This would be particularly useful for constructing a real-time disease warning system because we can quickly conduct inference when new confirmed cases are collected. 

In a Bayesian framework, it is natural to use previous window's information as a prior for current model fitting. The procedure of rolling-window updates are as follows. For a given time window $w$, we use a prior $\pi^{(w)}(\bm{\theta})$ as $N(\bm{\mu}^{(w-1)},c\Sigma^{(w-1)})\mathbf{I}(\bm{\theta}>0)$, where $\bm{\mu}^{(w-1)}$ is a posterior mean vector and $\Sigma^{(w-1)}$ is a diagonal matrix whose elements are the posterior variance of  parameters from window $w-1$. Here, we use the truncated normal prior to ensure range parameters are well defined. We multiply a constant $c=2$ to $\Sigma^{(w-1)}$ to make prior wider. Furthermore, in our MCMC updates, we use initial locations of cluster centers $\mathbf{c}^{(w)}$ from the last updated location of cluster centers $\mathbf{c}^{(w-1)}$. For $w=1$, we use a positive non-informative prior (i.e., $\pi^{(1)}(\bm{\theta})=\mathbf{I}(\bm{\theta}>0)$) and use randomly generated locations from landmarks as $\mathbf{c}^{(1)}$.

\begin{figure}[htbp]
    \begin{subfigure}{0.5\textwidth}
        \centering
        \includegraphics[width=\linewidth]{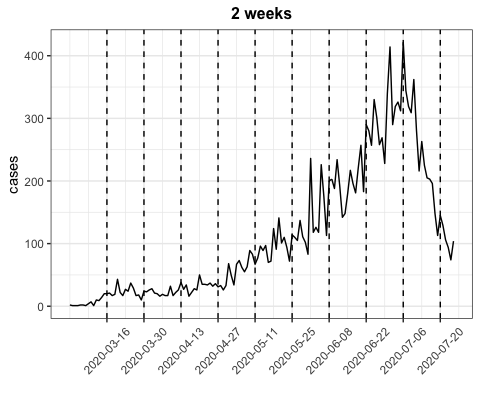}
    \end{subfigure}
        \begin{subfigure}{0.5\textwidth}
        \centering
        \includegraphics[width=\linewidth]{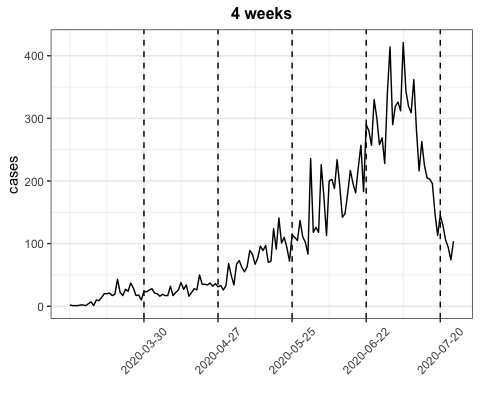}
    \end{subfigure}
    \caption{Confirmed COVID-19 cases of Cali, Colombia. Dashed lines indicate window borders.}
    \label{fig:timewindow}
\end{figure}

In this paper, we fit STDPG with two different windows (last two weeks and last four weeks) at any given time point. Figure~\ref{fig:timewindow} shows a time series plot of confirmed cases of COVID-19 with different time windows. The main goal of this study is to detect space-time cluster centers of confirmed cases as well as to examine the time-varying effects of parameters across different times of the disease spread. 

\subsection{Model assessment}

To validate our models, we compare the observed proportion of data points and the theoretical proportion of density from the likelihood function. First, we construct $G=8 \times 13 \times 3$ number of rectangular cubes covering our study domain $\mathcal{S}\times \mathcal{T}$. From \eqref{likelihood}, we can obtain the theoretical proportion of density over a cube $[a_g,b_g]^3$ as
\[
p^{t}_{g} = \int_{[a_g,b_g]^{3} \in \mathcal{S}\times T} \sum_{\forall j~\mbox{non-empty}}q_j \frac{1}{\omega_{s}^{2}\omega_{t}} \exp\Big(-\frac{||\mathbf{x}-\mathbf{c}^{s}_{j}||^2}{2\omega_{s}^{2}}-\frac{(t-c^{t}_{j})^2}{2\omega_{t}^{2}}\Big) d\mathbf{X}
\]
where $\mathbf{X} \in  \mathcal{S}\times T$ is the spatio-temporal point process. In practice, we obtain $\widehat{p^{t}_{g}}$ via numerical integration using ${\tt{pmvnorm}}$ function in {\tt R}. Similarly, we can calculate the observed proportion of data points located in a cube $[a_g,b_g]^3$ as 
\[
p^{o}_{g} = \frac{\mbox{the number of data points in}~[a_g,b_g]}{\mbox{total number of observations}}.
\]
Then, we draw Q-Q plots using $p_g^o,\widehat{p^{t}_{g}}$ calculated from $g=1,\cdots,G$. If our model fits well, Q-Q plots would follow a straight line. Furthermore, we compute  the mean-squared error between the observed and theoretical proportions as
\[
\mbox{MSE} = \frac{1}{G}\sum_{\forall g}(p^{t}_{g} - p^{o}_{g})^2.
\]
Naturally, the smaller MSE values are, the better the model fits.

%\textcolor{red}{chisquare-test?}
%\[
%\sum_{\forall g}\frac{(p^{t}_{g}-p^{o}_{g})^2}{p^{t}_{g}}
%\]
%degree of freedom is $G-1$ ($G$ is the number of cubes)

%https://courses.lumenlearning.com/odessa-introstats1-1/chapter/goodness-of-fit-test/#:~:text=The%20number%20of%20degrees%20of,of%20the%20chi%2Dsquare%20curve.

\section{Application}\label{sec:application}

Here, we apply our method to COVID-19 cases in Cali, Colombia. The proposed method is implemented through {\tt R} and {\tt C++} using the Rcpp and RcppArmadillo packages \citep{eddelbuettel2011rcpp}. The source code and data can be downloaded from  https://github.com/jwpark88/STDP. For each time window, we run the MCMC algorithm for 20,000 iterations, with the first 10,000 iterations discarded for burn-in. In our analysis, we set the maximum number of clusters $M=120$ for the two weeks window and $M=200$ for the four weeks window, which can ensure the number of non-empty clusters $M^{\ast}$ is smaller than $M$. For parameter interpretation, we convert spatial range parameters $\omega_s, \lbrace \omega_l \rbrace_{l=1}^{p}$ into \textit{kilometers} scale using the Haversine formula \citep{van2012heavenly} (see the supplementary material for details). We also convert the scale of temporal range parameter $\omega_t$ into \textit{days} by multiplying it by 28.

\begin{figure}[htbp]
\begin{center}
\includegraphics[width=0.8\linewidth]{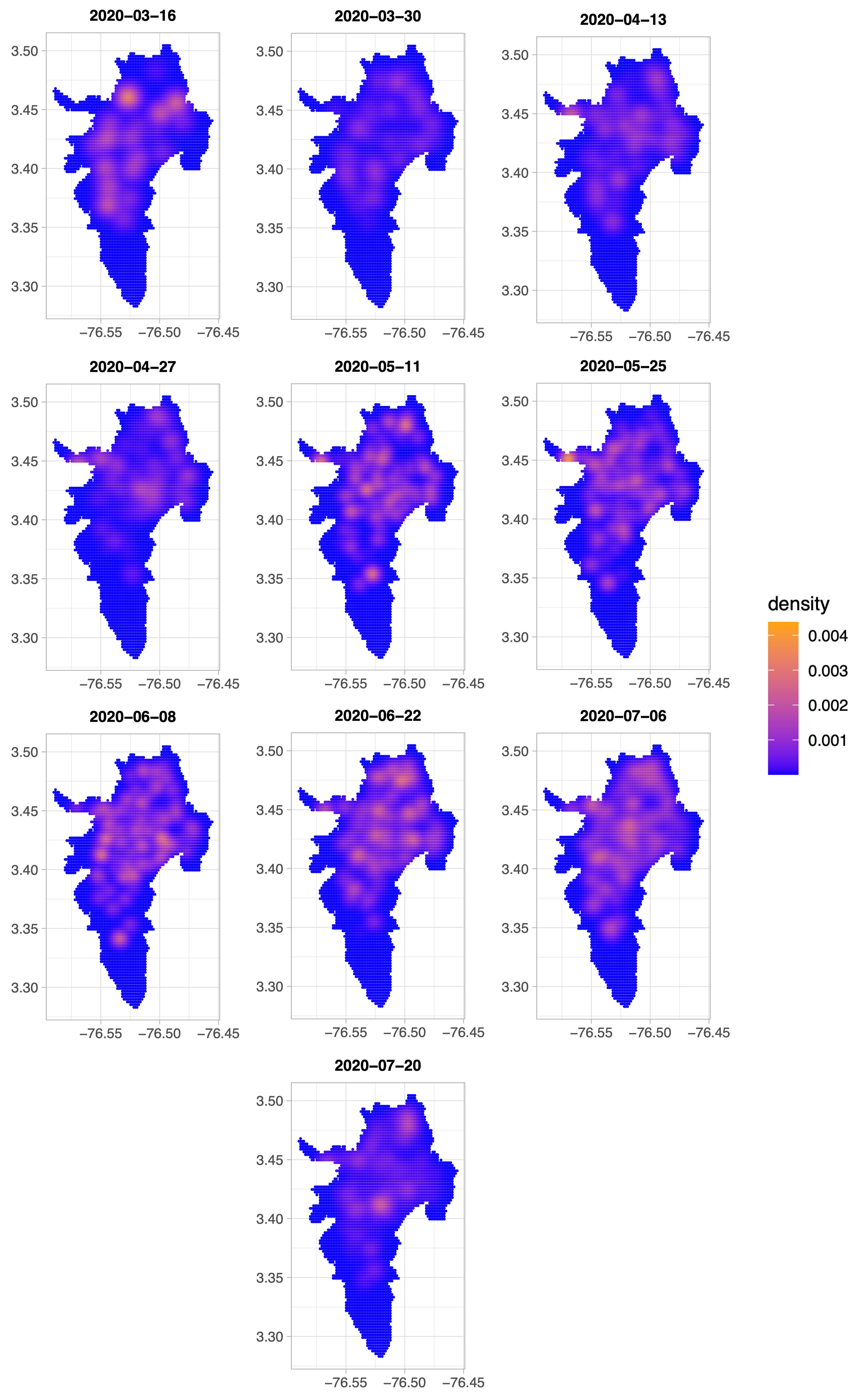}
\end{center}
\caption[]{Posterior mean of density map for the observed cases with two weeks intervals.}
\label{fig:density_2weeks}
\end{figure}

\begin{figure}[htbp]
\begin{center}
\includegraphics[width=0.9\linewidth]{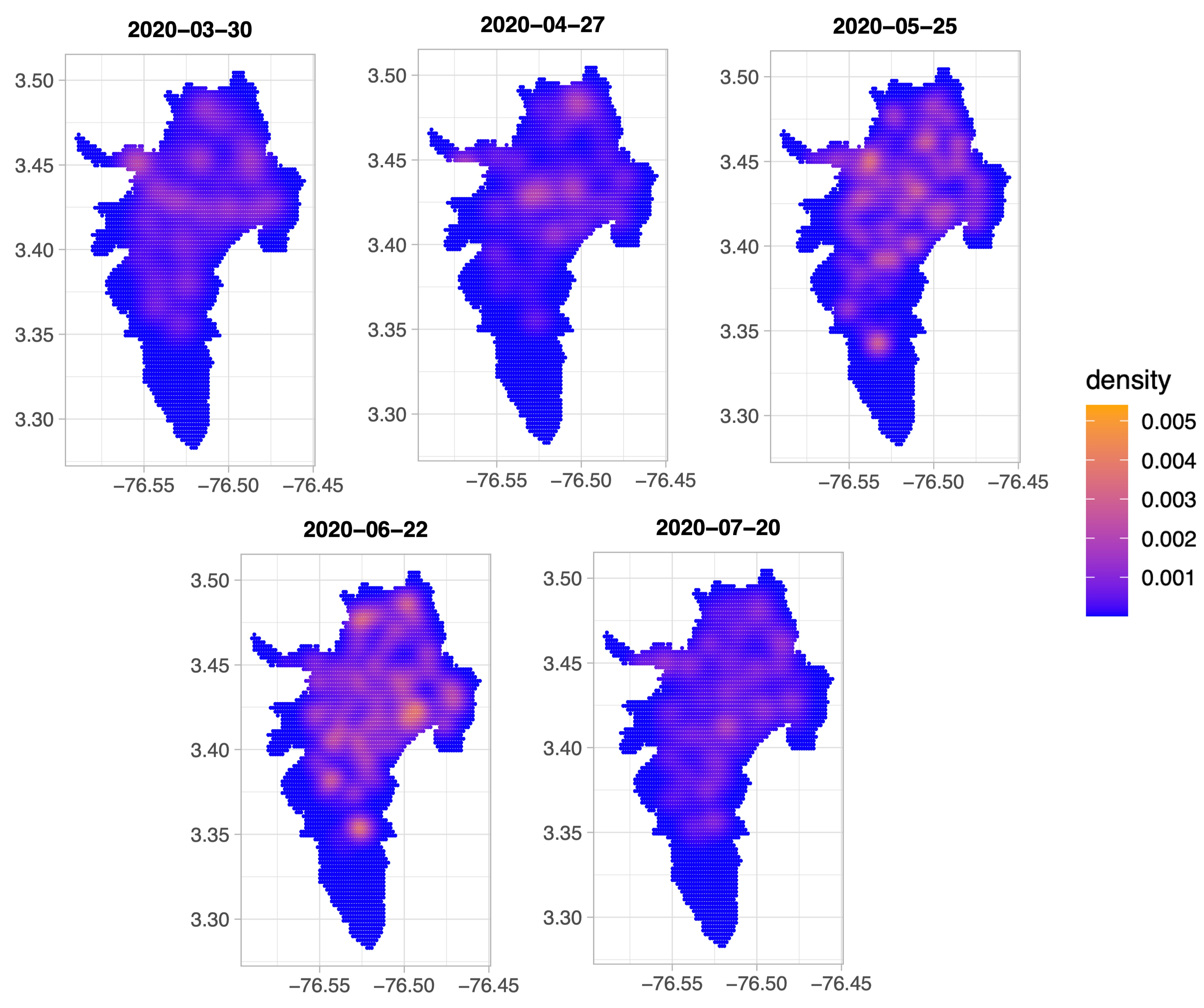}
\end{center}
\caption[]{Posterior mean of density map for the observed cases with four weeks intervals.}
\label{fig:density_4weeks}
\end{figure}

\begin{figure}[htbp]
    \begin{subfigure}{0.45\columnwidth}
        \centering
        \includegraphics[width=\linewidth]{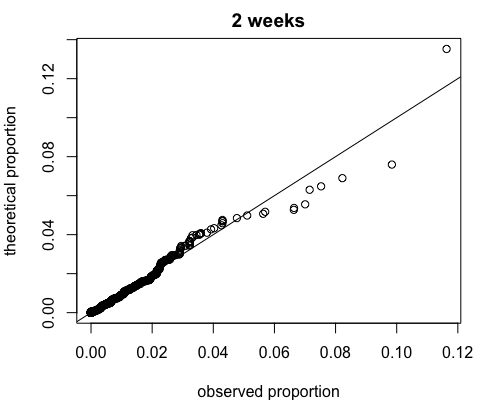}
        \caption{MSE=0.00004157}
        %\label{fig:my_label}
    \end{subfigure}
        \begin{subfigure}{0.45\columnwidth}
        \centering
        \includegraphics[width=\linewidth]{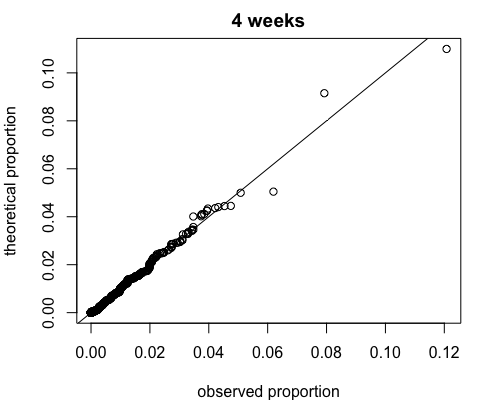}
        \caption{MSE=0.00001866}
        %\label{fig:my_label}
    \end{subfigure}
    \caption{Q-Q plots for $p_g^o$ and $\hat{p}_g$.}
    \label{fig:qqplot}
\end{figure}

\paragraph{Visualization of COVID-19 hot spots.} Figures \ref{fig:density_2weeks} (with two weeks intervals) and \ref{fig:density_4weeks} (with four weeks intervals) show the posterior mean surfaces of the spatio-temporal densities for the observed cases. They show how the detected hot spots vary and evolve with time, and
%clearly they differ depending on the past information we consider in feeding our model. 
both density maps are sort of similar indicating the way the disease was spreading. It started in the central and northeastern part of Cali, then showed generalized outbreaks later in May and June with increasing densities. Overall infected cases decreased later in July. 
%It started in the north eatern part of Cali, then several outbreaks started in the western part to show generalized outbreaks later in May and June in many central to northern areas of the city. 
To show how well these densities describe the spatio-temporal distribution of the observed cases, we created Q-Q plots for $p_g^o$ and $\widehat{p^{t}_{g}}$ as previously described. The plots in Figure \ref{fig:qqplot} show that the estimated spatio-temporal densities well capture the observed cases, as the estimated proportion $\widehat{p^{t}_{g}}$ is in good agreement with the observed proportion $p_g^o$ in both rolling windows.

\begin{figure}[htbp]
    \centering
    \includegraphics[width=.9\linewidth]{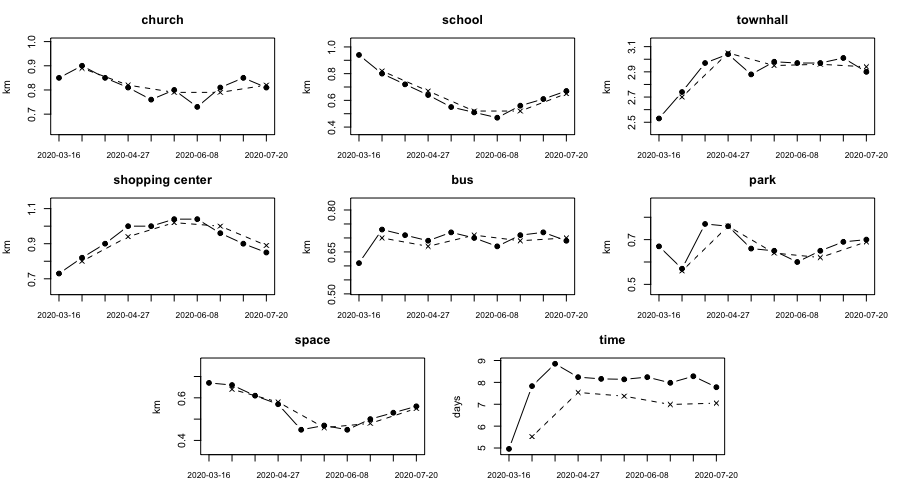}
    \caption{Posterior mean of range parameters across different times. Dashed (solid) lines indicate results from two weeks (four weeks).}
    \label{fig:mean_trend}
\end{figure}
%associated to the landmarks

\paragraph{Interpretation of landmark parameters.} To describe how the properties of case clusters change over time in detail, we plot the time series of posterior means of the range parameters ($\omega_s, \omega_t, \lbrace \omega_l \rbrace_{l=1}^{p}$)   in Figure \ref{fig:mean_trend}. We also provide posterior means and 95\% HPD intervals for all types of range parameters in the supplementary material; see Tables 1-4. As discussed in Section \ref{sec:model_framework}, a smaller range parameter value for a certain type of landmark means the clusters tend to show up closer to those landmarks. In other words, a smaller range value can be interpreted as a stronger influence of a landmark on the disease spread for the given period, whereas a larger value indicates a weaker influence.

The range parameter for school decreases as the cases increase for the first 14 weeks and then slightly increases as the cases decreases for the last 6 weeks. This indicates that schools may be more important in determining the disease hot spots when the number of cases is higher. The range parameter for the shopping center, on the other hand, moves in the opposite direction, showing that shopping centers play less important roles in detecting the hot spots when the cases are more prevalent. This is perhaps because the higher number of COVID-19 cases did not much discourage people to visit schools, while certainly discouraged people to go to shopping centers.  This suggests that the disease control effort should focus more on schools when the prevalence of disease is high, and more on shopping centers when the prevalence is low. 

The range parameter for churches and bus stations did not change much over the study period, being the values always quite low (around 0.8 for churches and 0.7 for bus stations). This implies that public transportation is always important in detecting hot spots regardless of the phase of the disease spread. The range parameter for parks shows an interesting pattern. At the beginning, parks seemed to play an important role in detecting disease clusters, and then they became less important in the following weeks, becoming more important again in the last 12 weeks. This shows that people decided to avoid parks on the onset of the pandemic, but then changed their behavior later and visited parks again. This is perhaps not surprising given that outdoor activities were viewed as a safer alternative to indoor activities during the pandemic period. 

The range parameter for the town hall shows much larger values, showing that the town hall is much less important in hot spot detection than the other types of landmarks in general. Its value was smaller in the first four weeks, and then became larger for the rest of the study period. This perhaps indicates that people decided to avoid the town center after seeing the rapid increase of the number of cases at the beginning, and then the behavior did not change even after the cases has decreased.

\begin{figure}[htbp]
\begin{center}
\includegraphics[height=7.5in]{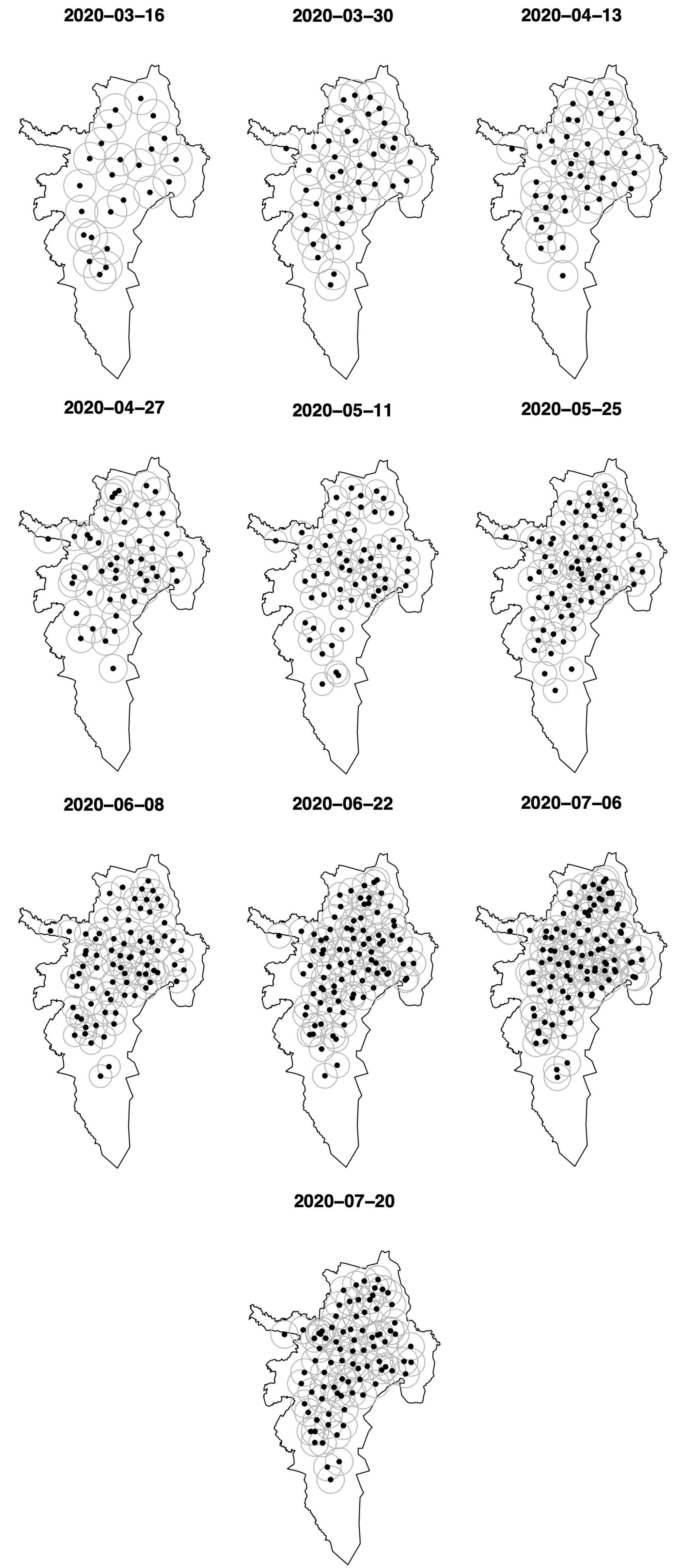}
\end{center}
\caption[]{Illustration of spatio-temporal clusters with two weeks intervals. Black dots indicate the location of cluster centers, and grey circles indicate their risk boundaries. The number of clusters (in order of time) is estimated 24, 41, 42, 49, 57, 71, 78, 89, 102, 86, respectively.}
\label{fig:cluster2}
\end{figure}

\begin{figure}[ht]
\begin{center}
\includegraphics[height=4.5in]{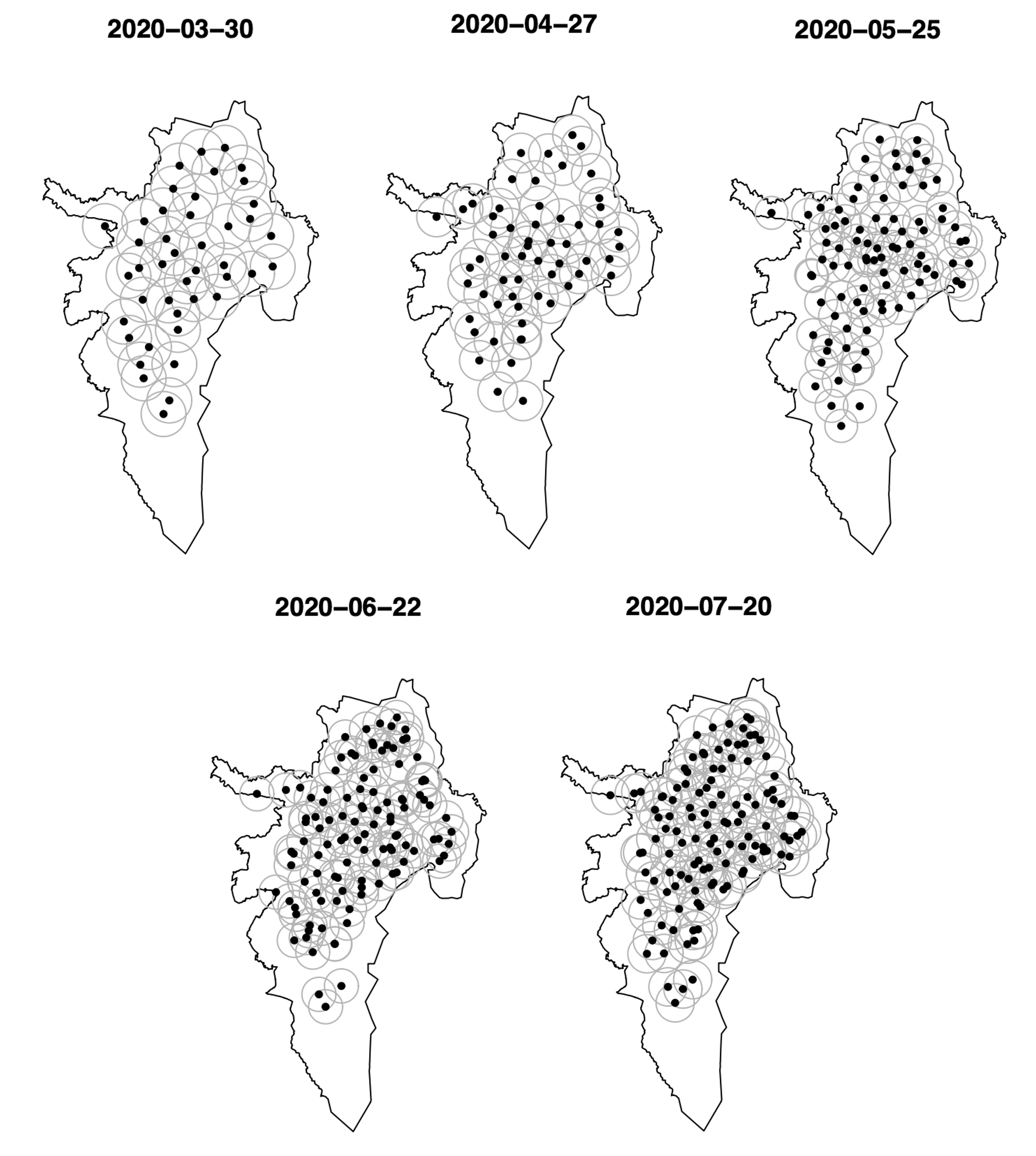}
\end{center}
\caption[]{Illustration of spatio-temporal clusters with four weeks intervals. Black dots indicate the location of cluster centers, and grey circles indicate their risk boundaries. The number of clusters (in order of time) is estimated 43, 59, 88, 112, 120, respectively.}
\label{fig:cluster4}
\end{figure}

\paragraph{Interpretation of space and time range parameters.} The range parameter for space takes values between 0.4$km$ and 0.7$km$ (Figure \ref{fig:mean_trend}), meaning about 95\% of the observed points are located within $2\times0.5=1km$ radius from the cluster center on average, because normal kernels are used. The range parameter for time starts with a value of 5 days and then increases to around 8 days in the two weeks window results, and starts with a value of 6 days and then increases to around 7 days in the four weeks window results (Figure \ref{fig:mean_trend}). This means that the 95\% of the observed values are within a two weeks range or slightly longer from the cluster center. 

Our approach can also visualize "risk boundaries" from parameter estimates; such boundaries would be useful for diagnosing the degree of COVID-19 risk. Figures \ref{fig:cluster2}, \ref{fig:cluster4} illustrate cluster centers and their spatial risk boundaries. Specifically, cluster centers $\mathbf{c}$ are sampled from the conditional posterior distribution (see supplementary material for details), and the radii are obtained by $2 \times \omega_s$. The range parameter for space seems to be negatively correlated to the number of total cases. As shown in Figures \ref{fig:cluster2} and \ref{fig:cluster4},  the number of kernels got larger and their sizes became smaller as the number of cases increased. This pattern coincides with what has been observed by \cite{park2022interaction} in their similar but simpler purely spatial model. Individual clusters became smaller as the spatial distribution of events became denser, which leads to a smaller value of the range parameter for spatial kernels. The range parameter for time shows a small discrepancy between the results based on two weeks intervals and four weeks intervals. The slightly shorter range (hence the longer time dependence) in the four weeks interval results seems to be a consequence of stronger smoothing effects due to the longer time window.

\section{Discussion}\label{sec:discussion}

In this manuscript, we have proposed a spatio-temporal point process method based on the Dirichlet process Gaussian mixture. Our model can detect spatio-temporal clusters of disease with pragmatic parameter interpretations, which is one of the advantages over existing mechanistic point process model frameworks \citep{diggle2013spatial, gonzalez2016spatio}. To provide a realistic and on-time inference, we have provided a rolling-window scheme by incorporating prior information from the posteriors from the previous time window. We have also provided a spatio-temporal density map from the fitted model, which gives an intuitive epidemiological interpretation. We have validated our model by comparing the expected densities and the observations. The results indicate that our model captures well the spatio-temporal pattern of observed events. The ideas and method proposed in this paper can generally be applicable to model other clustering processes such as distribution of rat sightings \citep{tamayo2014modelling} and dynamics of crime \citep{reinhart2018self}.

From a computational perspective, the use of Dirichlet process allows us to detect disease spreading centers quickly with much lower computational cost, compared to the Neyman-Scott process approaches \citep{mrkvivcka2014two,park2022interaction}; fitting the Neyman-Scott process model is computationally prohibitive for complex spatio-temporal processes due to the slow mixing of birth-death MCMC. In terms of modeling and application aspects, our method provides insights on disease hot spots, which are useful for planning public health policy. For example, posterior means of $\omega_s, \omega_t$ provide the spatio-temporal range of disease spreading hot spots. From $\lbrace \omega_l \rbrace$, we can estimate the impact of major landmarks during disease outbreaks. 

As per the future directions, one might consider adding interactions among spatio-temporal clusters through more advanced mixture models. For example, one might be able to formulate a model with a two-level hierarchy of clusters to introduce clustering behaviors of clusters.
Another possibility is to construct a dependent covariance structure between Gaussian mixture components at the expense of computational costs. Developing such extensions of our model to capture more complicated dependent events may provide an interesting avenue for future research. 

\section*{Supplementary Material}
The supplementary material available online provides details of the full conditionals for MCMC, Haversine formula, and additional tables.

\section*{Acknowledgement}
Jaewoo Park and Seorim Yi were supported by the National Research Foundation of Korea (NRF-2020R1C1C1A0100386813). Jorge Mateu was partially supported by the National project with reference PID2019-107392RB-I00/AEI/10.13039/501100011033.

%\nocite{*}
\bibliography{Reference.bib}

\clearpage
\appendix
\begin{center}
\title{\LARGE\bf Supplementary Material for A Spatio-Temporal Dirichlet Process Mixture Model for Coronavirus Disease-19}\\~\\
\author{\Large{Jaewoo Park, Seorim Yi, Won Chang, and Jorge Mateu}}
\end{center}

%---------------------------------------------------------------
\section{Full conditionals for Markov Chain Monte Carlo}\label{appA}

\begin{itemize}
\item 
The conditional distribution of range parameters $(\omega_s,\omega_t,\lbrace \omega_l \rbrace_{l=1}^{p})$ takes the form: 
\begin{align} 
\pi(\omega_s,\omega_t,\lbrace \omega_l \rbrace_{l=1}^{p}|\mbox{others}) & \propto  \prod_{i=1}^{N}f(\lbrace \mathbf{x}_i,t_i \rbrace |\omega_s,\omega_t,\lbrace \omega_l \rbrace_{l=1}^{p})\pi(\omega_s)\pi(\omega_t)\pi(\omega_1)\cdots\pi(\omega_p)\nonumber \\
& \propto \prod_{i=1}^{N}\Bigg[
\sum_{j=1}^{M}q_j \frac{1}{\omega_{s}^{2}\omega_{t}} \exp\Big(-\frac{||\mathbf{x}_i-\mathbf{c}^{s}_{j}||^2}{2\omega_{s}^{2}}-\frac{(t_i-c^{t}_{j})^2}{2\omega_{t}^{2}}\Big)\times \prod_{l=1}^{p}\frac{1}{\omega_{l}} \exp\Big(-\frac{||\mathbf{z}^{l}_{i}-\mathbf{c}^{s}_{j}||^2}{2\omega_{l}^{2}}\Big)\Bigg]\nonumber\\
& \times \pi(\omega_s)\pi(\omega_t)\pi(\omega_1)\cdots\pi(\omega_p)\nonumber
\end{align}

\item The conditional distribution of membership variable $g_i$ is of the form:
\begin{align} 
\pi(g_i|\mbox{others}) &= \mbox{Categorical}(\tilde{q}_{i1},\cdots,\tilde{q}_{iM}),~\mbox{where}  \nonumber\\ 
\tilde{q}_{ij} & \propto q_j \frac{1}{\omega_{s}^{2}\omega_{t}} \exp\Big(-\frac{||\mathbf{x}_i-\mathbf{c}^{s}_{j}||^2}{2\omega_{s}^{2}}-\frac{(t_i-c^{t}_{j})^2}{2\omega_{t}^{2}}\Big)\times \prod_{l=1}^{p}\frac{1}{\omega_{l}} \exp\Big(-\frac{||\mathbf{z}^{l}_{i}-\mathbf{c}^{s}_{j}||^2}{2\omega_{l}^{2}}\Big)\nonumber
\end{align}
for $j=1,\cdots,M$

\item The conditional distribution of space-time cluster centers $(\mathbf{c}^{s}_{j},c^{t}_{j})$ becomes:
\[
\pi((\mathbf{c}^{s}_{j},c^{t}_{j})|\mbox{others}) \propto \prod_{\forall i|g_i=j}\Bigg[
q_j \frac{1}{\omega_{s}^{2}\omega_{t}} \exp\Big(-\frac{||\mathbf{x}_i-\mathbf{c}^{s}_{j}||^2}{2\omega_{s}^{2}}-\frac{(t_i-c^{t}_{j})^2}{2\omega_{t}^{2}}\Big)\times \prod_{l=1}^{p}\frac{1}{\omega_{l}} \exp\Big(-\frac{||\mathbf{z}^{l}_{i}-\mathbf{c}^{s}_{j}||^2}{2\omega_{l}^{2}}\Big)\Bigg] \frac{1}{|\mathcal{S}|T}
\]
for $j=1,\cdots,M$
\item The conditional distribution of $U_j$:
\[
\pi(U_j|\mbox{others})= \mbox{Beta}(1+\sum_{i=1}^{N}I(g_i=j), b_u + \sum_{i=1}^{N}I(g_i>j))
\]

\item The conditional distribution of $b_u$ can be written as:
\[
\pi(b_u|\mbox{others})= \mbox{Gamma}(M-1+a, b-\sum_{j=1}^{M-1}\log(1-U_j)),
\]
where $a=1,b=1/4$
\end{itemize}

%---------------------------------------------------------------

\section{Haversine formula}

Consider latitude and longitude coordinates $(x_1,y_1), (x_2,y_2)$. Then we can calculate the great-circle distance $d$ (\textit{km}) between two coordinates using the Haversine formula, as follows: 
\[
dx = \frac{\pi(x_2-x_1)}{180},~~~dy= \frac{\pi(y_2-y_1)}{180}
\]
\[
a = \sin^2\Big(\frac{dx}{2}\Big) + \cos\Big( \frac{\pi x_1}{180} \Big) \cos\Big(\frac{\pi x_2}{180}\Big)\sin^2\Big(\frac{dy}{2}\Big)
\]
\[
d= 6371 \times 2 \arctan2(\sqrt{a},\sqrt{1-a}).
\]

\section{Additional Tables}

\begin{table}[htbp]
\centering
\begin{tabular}{ccc}
\hline
%March 19th 
  & space & time \\
  \hline
3/16 &  0.67 & 4.96 \\
     &  (0.59, 0.74) & (4.46, 5.54)\\
3/30 &  0.66 & 7.84\\
     &  (0.61, 0.7) & (7.36, 8.36) \\
4/13 &  0.61 & 8.84\\
     &  (0.57, 0.65) & (8.28, 9.48)\\
4/27 &  0.57 & 8.24\\
     &  (0.54, 0.6) & (7.68, 8.78)\\
5/11 &  0.45 & 8.16\\
     &  (0.43, 0.47) & (7.74, 8.56)\\
5/25 &  0.47 & 8.14\\
     &  (0.46, 0.48) & (7.8, 8.48)\\
6/8  &  0.45 & 8.24\\
     &  (0.44, 0.46) & (7.98, 8.5)\\
6/22 &  0.5 & 7.98\\
     &  (0.49, 0.51) & (7.74, 8.18)\\
7/6  &  0.53 & 8.28\\
     &  (0.52, 0.54) & (8.04, 8.5)\\
7/20 &  0.56 & 7.78\\
     &  (0.55, 0.57) & (7.54, 8)\\
\hline
\end{tabular}
\caption{Posterior means and 95\% HPD intervals (parenthesis) of space, time parameters for two weeks window.}
\label{2week} 
\end{table}

\begin{table}[htbp]
\centering
\begin{tabular}{ccccccc}
\hline
%March 19th 
  & church & school & town hall & shopping center & bus & park \\
  \hline
3/16 &  0.85 & 0.94 & 2.53 & 0.73 & 0.61 & 0.67\\
     & (0.75, 0.94) & (0.85, 1.04) & (2.29, 2.77) & (0.64, 0.81) & (0.53, 0.7) & (0.59, 0.77)\\
3/30 & 0.9 & 0.8 & 2.74 & 0.82 & 0.73 & 0.57 \\
     & (0.86, 0.96) & (0.76, 0.85) & (2.6, 2.89) & (0.77, 0.87) & (0.69, 0.77) & (0.53, 0.62) \\
4/13 & 0.85 & 0.72 & 2.97 & 0.9 & 0.71 & 0.77 \\
     & (0.8, 0.89) & (0.67, 0.76) & (2.84, 3.1) & (0.86, 0.95) & (0.67, 0.75) & (0.71, 0.82) \\
4/27 & 0.81 & 0.64 & 3.04 & 1 & 0.69 & 0.76\\
     & (0.78, 0.85) & (0.61, 0.68) & (2.92, 3.16) & (0.95, 1.05) & (0.65, 0.72) & (0.72, 0.8) \\
5/11 & 0.76 & 0.55 & 2.88 & 1 & 0.72 & 0.66\\
     & (0.74, 0.79) & (0.53, 0.57) & (2.78, 2.97) & (0.96, 1.03) & (0.69, 0.74) & (0.63, 0.68)\\
5/25 & 0.8 & 0.51 & 2.98 & 1.04 & 0.7 & 0.65\\
     & (0.78, 0.82) & (0.49, 0.53) & (2.91, 3.05) & (1.01, 1.06) & (0.68, 0.73) & (0.63, 0.67)\\
6/8 & 0.73 & 0.47 & 2.97 & 1.04 & 0.67 & 0.6\\
     & (0.72, 0.75) & (0.46, 0.49) & (2.9, 3.03) & (1.02, 1.06) & (0.65, 0.68) & (0.58, 0.61)\\
6/22 & 0.81 & 0.56 & 2.97 & 0.96 & 0.71 & 0.65\\
     & (0.8, 0.82) & (0.55, 0.57) & (2.92, 3.02) & (0.95, 0.98) & (0.69, 0.72) & (0.64, 0.67)\\
7/6 & 0.85 & 0.61 & 3.01 & 0.9 & 0.72 & 0.69\\
     & (0.83, 0.86) & (0.6, 0.62) & (2.97, 3.05) & (0.89, 0.91) & (0.71, 0.73) & (0.68, 0.7) \\
7/20 & 0.81 & 0.67 & 2.9 & 0.85 & 0.69 & 0.7\\
     & (0.8, 0.82) & (0.66, 0.68) & (2.86, 2.94) & (0.84, 0.87) & (0.67, 0.7) & (0.69, 0.71)\\
\hline
\end{tabular}
\caption{Posterior means and 95\% HPD intervals (parenthesis) of landmark parameters for two weeks window.}
\label{2weeklandmark} 
\end{table}

\begin{table}[htbp]
\centering
\begin{tabular}{ccc}
\hline
%March 19th 
  & space & time \\
  \hline
3/30 & 0.64 & 5.52\\
     & (0.61, 0.68) & (5.13, 5.95)\\
4/27 & 0.58 & 7.54\\
     & (0.56, 0.6) & (7.18, 7.9)\\
5/25 & 0.46 & 7.37\\
     & (0.45, 0.47) & (7.14, 7.6)\\
6/22 & 0.48 & 6.99\\
     & (0.47, 0.48) & (6.78, 7.21)\\
7/20 & 0.55 & 7.05\\
     & (0.54, 0.56) & (6.92, 7.19)\\
\hline
\end{tabular}
\caption{Posterior means and 95\% HPD intervals (parenthesis) of space, time parameters for four weeks window.}
\label{4weekst} 
\end{table}

\begin{table}[htbp]
\centering
\begin{tabular}{ccccccc}
\hline
%March 19th 
  & church & school & town hall & shopping center & bus & park \\
  \hline
3/30 & 0.89 & 0.82 & 2.7 & 0.8 & 0.7 & 0.56\\
     & (0.84, 0.93) & (0.78, 0.87) & (2.59, 2.84) & (0.76, 0.85) & (0.66, 0.75) & (0.53, 0.6)\\
4/27 & 0.82 & 0.67 & 3.05 & 0.94 & 0.67 & 0.76\\
     & (0.78, 0.85) & (0.64, 0.7) & (2.94, 3.15) & (0.91, 0.97) & (0.65, 0.71) & (0.74, 0.79)\\
5/25 & 0.79 & 0.52 & 2.95 & 1.02 & 0.71 & 0.64\\
     & (0.77, 0.8) & (0.51, 0.55) & (2.9, 3.01) & (1, 1.04) & (0.69, 0.72) & (0.63, 0.66)\\
6/22 & 0.79 & 0.52 & 2.96 & 1 & 0.69 & 0.62\\
     & (0.77, 0.8) & (0.51, 0.53) & (2.92, 3) & (0.98, 1.01) & (0.68, 0.71) & (0.61, 0.63)\\
7/20 & 0.82 & 0.65 & 2.94 & 0.89 & 0.7 & 0.69\\
     & (0.81, 0.83) & (0.64, 0.65) & (2.91, 2.98) & (0.87, 0.9) & (0.69, 0.71) & (0.68, 0.7)\\
\hline
\end{tabular}
\caption{Posterior means and 95\% HPD intervals (parenthesis) of landmark parameters for four weeks window.}
\label{4weeklandmark} 
\end{table}

\end{document}